\documentclass[graybox]{svmult}

\usepackage{type1cm}        
\usepackage{makeidx}         
\usepackage{graphicx}        
\usepackage{multicol}        
\usepackage[bottom]{footmisc}
\usepackage{booktabs}
\usepackage{amsmath}
\usepackage{tabularx}
\usepackage{float}

\usepackage{hyperref}

\usepackage{newtxtext}       %
\usepackage[varvw]{newtxmath}       

\makeindex             

\begin{document}

\title*{The accretion, emission, mass and spin evolution of primordial black holes}
\author{Valerio De Luca and Nicola Bellomo}
\institute{Valerio De Luca \at Center for Particle Cosmology, Department of Physics and Astronomy,
University of Pennsylvania 209 S. 33rd St., Philadelphia, PA 19104, USA. 
\email{vdeluca@sas.upenn.edu}
\and Nicola Bellomo \at Texas Center for Cosmology and Astroparticle Physics, Weinberg Institute, Department of Physics, The University of Texas at Austin, Austin, TX 78712, USA.\\
Dipartimento di Fisica e Astronomia G. Galilei, Universit\`{a} degli Studi di Padova, Via Marzolo 8, 35131 Padova, Italy.
\email{nicola.bellomo@unipd.it}}
\maketitle

\abstract{Throughout the cosmic history, primordial black holes may experience an efficient phase of baryonic mass accretion from the surrounding medium. While the realm of accretion physics is marked by numerous uncertainties, and a comprehensive understanding remains elusive, recent investigations have delved into this area, exploring its implications for the cosmological evolution of these compact objects. Notably, primordial black holes could experience characteristic growths of their masses and spins, accompanied by the emission of radiation,  ultimately responsible for feedback effects that could weaken the efficiency of the process. In this chapter we review the basic formalism to describe the accretion rate, luminosity function and feedback effects, in order to provide distinctive predictions for the evolution of the primordial black hole mass and spin parameters.}

\section{Accretion model} 
\label{Sec1}

During their cosmological evolution, primordial black holes (PBHs) can efficiently accrete particles from the medium through which they move. This process can happen either for isolated PBHs, which individually flow in the surrounding gas, or for PBHs which are assembled in binary systems from the epoch of matter-radiation equality, and it can occur both in the early stages of the universe or within the cosmological structures formed in the late universe. An efficient phase of accretion strongly impacts on the evolution of these cosmic relics by increasing their mass and spin parameters, as well as converting a fraction of the accreted mass to radiation. The injection of the latter in the primordial plasma could change the thermal and ionization histories, leading to distortions in the frequency spectrum and temperature/polarization power spectra of the CMB. The corresponding constraints on a PBH population undergoing an efficient accretion phase will be thoroughly discussed in Chapter 26.

The purpose of this Section is instead to summarise the main formalism to model accretion for individual or binary PBHs, in order to set the stage for computing the evolution of their mass and spin parameters. The interested reader can find more comprehensive discussions and complementary details in the following list of references~\cite{Ricotti:2007jk,Ricotti:2007au,Ali-Haimoud:2016mbv, DeLuca:2020bjf,DeLuca:2020qqa}.
In this Chapter we will adopt natural units $\hbar = c = 1$.

\subsection{Accretion onto individual PBHs}

During the early stages of the universe, individual PBHs orbiting in the primordial gas could accrete dark matter and baryons within the so-called Bondi radius, describing the distance in which the PBH gravitational potential dominates the dynamics of the infalling particles. The main effects of this process are the consequential increase of the PBH masses and spins, as well as the amount of radiation injected in the surrounding fluid. The latter may ionize and/or heat the accreting gas, affecting the radiative output and giving rise to feedback effects. A dedicated discussion of these effects is postponed to the next Section.

In the following we provide a general description of accretion, focusing on PBHs which are either in isolation ({\it naked}) or have an envelope of dark matter particles surrounding them ({\it dressed}). The latter situation may be realised when the bulk of the dark matter is not made of PBHs, but of a dominant component that seeds, after the redshift of matter-radiation equivalence, the growth of a dark halo. Its additional contribution to the PBH gravitational potential may thus increase the gas accretion rate onto the central PBH by several orders of magnitude, affecting their cosmological evolution. Finally, we introduce the concept of luminosity, in order to quantitatively estimate the amount of radiation emitted in the medium during the accretion process. 

\subsubsection{Basic formalism for spherical accretion}

We consider accretion of a hydrogen gas onto an isolated point mass, bathed in the quasi-uniform CMB radiation field. Aside from radiative feedback effects, which we neglect in the first place to simplify the picture, the second aspect to consider is 
the geometry of the accretion flow. In particular, if the characteristic angular momentum of the accreted gas is smaller than the angular momentum at the innermost stable circular orbit, then accretion has mostly a spherical flow. Otherwise, an accretion disk may form. The latter could easily catalyse the emission of radiation by bremsstrahlung from the hot ionized plasma near the event horizon, while in the spherical case the large viscous heating required to dissipate angular momentum leads to radiating a significant fraction of the rest-mass energy~\cite{1983bhwd.book.....S}.  Since the estimate of the gas angular momentum requires dedicated discussions on the PBH-baryon relative velocity, we first derive the most conservative accretion rate and luminosity by adopting a spherical accretion model, and postpone the discussion of the disk formation to the next Section.

In general, one should solve for the coupled, time-dependent, fluid, heat and ionization equations. However, as long as the characteristic accretion timescale is much shorter than the Hubble timescale, one can make the steady-state approximation and adopt a simplified Newtonian treatment in the region far from the BH horizon~\cite{Ali-Haimoud:2016mbv}.
The typical accretion time scale is given in terms of the Salpeter time by
\begin{equation}
	\tau_\text{\tiny acc} \equiv \frac{\tau_\text{\tiny Salp}}{\dot m}=\frac{\sigma_\text{\tiny T}}{4 \pi 
		m_p} \frac{1}{\dot m} =\frac{4.5 \times 10^8 \,{\rm yr}}{\dot m}\,, 
\label{tauACC}
\end{equation}
where $\sigma_\text{\tiny T}$ is the Thompson cross section, $m_p$ is the proton mass and $\dot m$ indicates the dimensionless accretion rate normalised to the Eddington one
\begin{equation}
\dot m = \frac{\dot{M}}{\dot{M}_\text{\tiny Edd}}
\quad\text{with}\quad 
\dot{M}_\text{\tiny Edd}= 1.44 \times 10^{17} \left( \frac{M}{M_\odot}\right) \rm{g \, s^{-1}}\,.
\label{mdot}
\end{equation}
The behaviour of the accretion rate $\dot{M}$ as a function of redshift $z$ and PBH mass $M$ will be the main focus of this Section. As showed in Refs.~\cite{Ricotti:2007jk,Ricotti:2007au}, for PBH masses $M \lesssim 10^{4}M_\odot$, a steady-state approximation is satisfied, so we shall limit ourselves to this mass range to simplify the time-dependent problem. 

The steady-state mass and momentum equations for the fluid are
\begin{align}
    4 \pi r^2 \rho |v| &= \dot{M} = \textrm{const}\,, \label{eq:mass} \\
    v \frac{d v}{d r} &= - \frac{G M}{r^2}  - \frac1{\rho} \frac{d P}{dr} - H v - \frac{4}{3} \frac{\overline{x}_e \sigma_\text{\tiny T} \rho_\text{\tiny CMB}}{m_p} v\,, \label{eq:momentum}
\end{align}
in terms of the peculiar radial velocity $v$ of the accreted gas (i.e. the velocity with respect to the Hubble flow), its density $\rho$, temperature $T$, and pressure $P$, given by
\begin{equation}
P = \frac{\rho}{m_p}(1 + \overline{x}_e) T\,,
\end{equation}
where the ionization fraction $x_e$ has been equated to its background value $\overline{x}_e$ in the absence of feedback effects~\cite{Ali-Haimoud:2016mbv}.
Following the steady-state approximation, we have neglected partial time derivative $\partial v/\partial t$, as well as the self-gravity of the accreted gas~\cite{Ricotti:2007jk}. Furthermore, the Hubble drag term $H v$ has the same form of the last term in the momentum equation, that describes the drag force due to Compton scattering of the ambient nearly homogeneous CMB photons with energy density $\rho_\text{\tiny CMB}$, which therefore contribute as an effective viscosity term in the equation~\cite{Ricotti:2007jk}.

The last equation of the system is the heat equation, that reads 
\begin{equation}
v \rho^{2/3} \frac{d}{dr}\left( \frac{T}{\rho^{2/3}}\right) = \frac{8 \overline{x}_e \sigma_\text{\tiny T}\rho_\text{\tiny CMB}}{3 m_e (1 + \overline{x}_e)}(T_\text{\tiny CMB} - T)\,, \label{eq:heat-compt}
\end{equation}
considering only Compton cooling by CMB photons  with temperature $T_\text{\tiny CMB}$ as a heat sink in this region~\cite{Peebles:1968ja}. In the steady-state approximation this equation enforces the condition $T (r \to \infty ) = T_\text{\tiny CMB}$. Similarly, at large distances from the PBH, the density should approach the baryonic one, $\rho (r \to \infty ) = \rho_\text{\tiny gas}$, with
\begin{equation}
\rho_\text{\tiny gas} =  250 \, \mu m_p \left( \frac{1+z}{1000} \right)^3\, {\rm g} \,  {\rm cm}^{-3}\,,
\end{equation}
for an hydrogen gas with mean molecular weight  $\mu = 1.22$.

The gas pressure and density far from the point mass allow to define an effective velocity $v_\text{\tiny eff} = \sqrt{P(r \to \infty)/\rho (r \to \infty)}$ (to be discussed below), such that an isolated PBH, moving with a relative velocity $v_\text{\tiny rel}$ with respect to the surrounding gas with sound speed $c_s$, accretes particles with a rate given by the Bondi-Hoyle expression~\cite{1983bhwd.book.....S}
\begin{equation}
\label{R1}
\dot{M}_\text{\tiny B} = 4 \pi \lambda \rho_\text{\tiny gas} v_\text{\tiny eff} r_\text{\tiny B}^2\,,
\end{equation}
where we have defined the Bondi radius as the region where accretion effectively occurs~\cite{1944MNRAS.104..273B}
\begin{equation}
r_\text{\tiny B} \equiv \frac{G M}{v_\text{\tiny eff}^2} \simeq 1.3 \times 10^{-4}\left( \frac{M}{M_\odot} \right) 
\left( \frac{v_\text{\tiny eff}}{5.7 {\rm\, km \, s^{-1}}} \right)^{-2}\,  {\rm pc}\,, 
\label{rBondi}
\end{equation}
and  $v_\text{\tiny eff} = \sqrt{v_\text{\tiny rel}^2 + c_s^2}$. For a gas in equilibrium at the temperature of the intergalactic medium, the corresponding sound speed can be approximated as~\cite{Ricotti:2007jk, Ricotti:2007au,DeLuca:2020bjf}
\begin{equation}
    c_s = \sqrt{\frac{k_B T_\text{\tiny gas}}{\mu m_p}} \simeq 5.7 \,  \left ( \frac{1+z}{1000}\right)^{1/2}
    \left[ \left (\frac{1+ z_\text{\tiny dec}}{1+z} \right)^\beta +1 \right]^{-1/2 \beta}\, {\rm km \, s^{-1}}\,,
\end{equation}
in terms of the Boltzmann constant $k_B$ and the gas temperature $T_\text{\tiny gas}$, that depends on the physics at large scales. In the second equality, by neglecting the feedback effect induced by accreting PBHs on the gas (to be discussed in the next Section), one can express it as a function of
the redshift  $z_\text{\tiny dec} \simeq 130 (\Omega_b h^2/0.022)^{2/5}$, at which the gas temperature deviates significantly from the CMB temperature, and of the characteristic exponent $\beta = 1.72$.

The constant term in the fluid equation, expressed by the accretion parameter $\lambda$, is introduced to capture, in a phenomenological manner, the influence of various non-gravitational forces that could oppose the gravitational pull of the compact object and thereby reduce the accretion rate.
For spherical accretion onto a point mass, it is of order unity and keeps into account the effects of the Hubble expansion, 
the coupling of the CMB radiation to the gas through Compton scattering, and the gas viscosity~\cite{Ricotti:2007jk}. Its analytical approximation is given by~\cite{Ali-Haimoud:2016mbv}
\begin{equation}
\lambda = \frac{1}{\lambda_\text{\tiny iso}}\left[\lambda_\text{\tiny ad}+(\lambda_\text{\tiny iso}-\lambda_\text{\tiny ad})\left(\frac{\gamma^{2}_{\rm c}}{88+\gamma^{2}_{\rm c}}\right)^{0.22}\right] \times  \left[\frac{1}{(\sqrt{1+\beta_{\rm c}}+1)^2}\exp\left(\frac{9/2}{3+\beta_{\rm c}^{3/4}}\right)\right],
\label{eq:lambda_nohalo}
\end{equation}
where $\lambda_\text{\tiny ad}=(3/5)^{3/2}/4\approx0.12$ and $\lambda_\text{\tiny iso}=e^{3/2}/4\approx1.12$. The coefficients 
$\beta_{\rm c}$ and $\gamma_{\rm c}$ are the dimensionless Compton drag and cooling rates from the surrounding CMB photons respectively, and they read~\cite{Ali-Haimoud:2016mbv}
\begin{eqnarray}
    \beta_{\rm c} = \frac{4}{3} \frac{x_{e}\sigma_\text{\tiny T}\rho_\text{\tiny gas}}{m_{p}}\frac{r_\text{\tiny B}}{v_\text{\tiny eff}}\,, \qquad 
    \gamma_{\rm c} = \frac{8}{3} \frac{x_{e}\sigma_\text{\tiny T}\rho_\text{\tiny gas}}{m_{e}(1+x_{e})}\frac{r_\text{\tiny B}}{v_\text{\tiny eff}}\,.
    \label{eq:gamma_c}
\end{eqnarray}
The knowledge of the exact value of $\lambda$ requires computing the solution to the  system of coupled equations. Some limits can however be deduced analytically. First,
when $\gamma_c \gg 1$, Compton cooling efficiently maintains thermal equilibrium down to distances smaller than the Bondi radius. At that point pressure forces are negligible relative to gravity, and the temperature is no longer relevant to the other fluid variables. For this isothermal setup, the expression simplifies to
 \begin{equation}
\lambda = \frac{1}{(\sqrt{1+\beta_{\rm c}}+1)^2} {\rm exp} \left( \frac{9/2}{3 + \beta_{\rm c}^{3/4}} \right)\quad  \text{for} \quad \gamma_c \gg 1\,.
\end{equation}
On the other hand, for negligible Compton drag $\beta_c \ll 1$, the momentum equation can no longer be rewritten as a conservation equation and one must solve explicitly the coupled fluid and heat equations, and determine the parameter $\lambda$ numerically. A good fit to the numerical results has been found in~\cite{Ali-Haimoud:2016mbv} and reads
\begin{equation}
\lambda \approx \lambda_\text{\tiny ad}+(\lambda_\text{\tiny iso}-\lambda_\text{\tiny ad})\left(\frac{\gamma^{2}_{\rm c}}{88+\gamma^{2}_{\rm c}}\right)^{0.22} \quad  \text{for} \quad \beta_c \ll 1\,.
\end{equation}
The combination of these two results allows for the approximated expression shown in Eq.~\eqref{eq:lambda_nohalo}, which is strictly valid for $\beta_c \ll \gamma_c$ and for $\beta_c \ll 1, \gamma_c \gg 1$.

The final ingredient in the expression of the accretion rate is the BH peculiar velocity $v_\text{\tiny rel}$, that has to be self-consistently accounted for to properly estimate the gas temperature and the dynamics of the accretion process.  While a dedicated time-dependent hydrodynamical simulation is probably needed to provide an exact result, given the dependence of the PBH proper motion on the amplitude of the inhomogeneities of the DM and baryon fluids, some analytical insights can be derived. 
Following Ref.~\cite{Ricotti:2007au}, and assuming PBHs to behave like DM particles, one can identify two main regimes of interest for the PBH relative velocity $v_\text{\tiny rel}$: a Gaussian linear contribution on large scales $v_\text{\tiny L}$, whose power spectrum and variance can be extracted from linear Boltzmann codes, and a small-scale contribution due to non-linear clustering of PBHs, $v_\text{\tiny NL}$. 

In the linear regime, gas and dark matter perturbations are coupled to each other until the decoupling redshift $z_\text{\tiny dec} \simeq 130 $ is reached. At earlier times,  
 the Silk damping acts suppressing the growth of inhomogeneities on small scales, such that the PBH peculiar velocity is of order of the gas sound speed. On the other hand, at redshifts  $z<z_\text{\tiny dec}$ (before entering the nonlinear regime), the gas flow lags behind the DM with a relative velocity $v_\text{\tiny L} = v_\text{\tiny DM}-v_{\text{\tiny b}}$. This velocity has a  complex time and scale-dependence, since baryons undergo acoustic oscillations and the dark-matter overdensities grow (see for example Fig.~2 of~\cite{Ricotti:2007au}). Ref.~\cite{Dvorkin:2013cea} explicitly compute $\langle v_\text{\tiny L}^2 \rangle$ as a function of time and find that it is mostly constant for $z \gtrsim 10^3$ and decays like $1+z$ at smaller redshifts. It therefore follows the simple dependence~\cite{Dvorkin:2013cea}
\begin{equation}
\sqrt{\langle v_\text{\tiny L}^2 \rangle}  \approx  \min\left[1, z/10^3\right] \times 30 ~\textrm{km/s}\,.
\label{eq:v_rms}
\end{equation}
As discussed above, the PBH  accretion rate depends on the combination $v_\text{\tiny eff} = \sqrt{v_\text{\tiny rel}^2 + c_s^2}$, such that the total energy injected in the plasma is obtained by averaging over the Gaussian distribution of relative velocities. The final average reads~\cite{Ricotti:2007au}
\begin{align}
	\langle v_\text{\tiny eff} \rangle_\text{\tiny A} &\sim c_s \left( \frac{16}{\sqrt{2\pi}} {\cal M}^3 \right)^{\frac{1}{6}} \theta 
({\cal M}-1) + c_s \left( 1 + {\cal M}^2 \right)^{\frac{1}{2}} \theta (1- {\cal M})\,,\nonumber \\
	\langle v_\text{\tiny eff} \rangle_\text{\tiny B} &\sim c_s {\cal M} \left[ \sqrt\frac{2}{\pi} {\rm ln}\left( \frac{2}{e}{\cal M} 
\right) \right]^{-\frac{1}{3}} \theta ({\cal M}-1) + c_s \left( 1 + {\cal M}^2 \right)^{\frac{1}{2}} \theta (1- {\cal M})\,,
	\end{align}
in terms of the Heaviside step function $\theta$ and Mach number ${\cal M} = \langle v_\text{\tiny rel} \rangle/c_s$, that identifies a supersonic $({\cal M} >1)$ or subsonic $({\cal M} <1)$ motion. Here the scenario A refers to a low efficient accretion rate, $\dot m < 1$, characterised by a spherical geometry, while scenario B refers  to an efficient accretion rate, $\dot m > 1$, which, as shown below, could support the presence of an accretion disk.

In the nonlinear regime, at redshifts $z \lesssim \mathcal{O}(10)$, the growth of nonlinear perturbations dominates the proper motion of PBHs. The velocity of PBHs which are massive enough to fall into the potential wells of the first cosmic structures could be sufficiently large to stop accretion, until they came to rest at the center of the host halo. At that point, it is unclear whether they meet favorable conditions for accretion to start again.
One can assume the characteristic velocity induced by nonlinear structures to be comparable with the circular velocity of virialized halos as a function of redshift. By determining their masses following the Press-Schechter approach, the circular and thus typical velocity reads~\cite{Ricotti:2007au}
\begin{equation}
v_\text{\tiny eff} \simeq (8 \cdot 10^2 \, {\rm km/s} ) \, \exp [-0.6 (1+z)] \sqrt{ \frac{1+z}{10} }\,.
\end{equation}
This estimate, based on a fluid-like approximation for the formation of dark matter halos, may suffer from few drawbacks: first, the discrete nature of PBHs, as well as their spatial distribution at formation, may change the power spectrum at small scales and invalidate the Press-Schechter assumption; second, this approach does not take into account kinetic effects, such as the thermal velocity distribution around the bulk motion velocity, and the putative presence of shocks and instabilities, which may hamper the applicability of the approach to small scales~\cite{Poulin:2017bwe}.
As further discussed in the next Section about feedback effects, the beginning of structure formation and the associated nonlinearities provide therefore a large source of uncertainty in the efficiency and duration of the accretion process.

The setup outline above describes naked PBHs, i.e. black holes in isolation in the surrounding medium. However, if PBHs do not comprise the totality of the dark matter $(f_\text{\tiny PBH} \lesssim 1)$, an additional and dominant dark matter component would be present around them and modify the accretion picture.
Current observational constraints show that PBHs with masses larger than ${\cal O}(M_\odot)$ can 
comprise only a fraction of the dark matter in the universe~\cite{Carr:2020gox}. In this case, PBHs would attract the surrounding dark matter particles,
generating an extended dark matter halo during the matter-dominated epoch with mass~\cite{Mack:2006gz, Adamek:2019gns, Berezinsky:2013fxa, Boudaud:2021irr}
\begin{equation}
\label{halo mass}
M_h(z) = 3 M \left( \frac{1+z}{1000} \right)^{-1}\,.
\end{equation}
It grows with time as long as PBHs are in isolation, and stops when all the available dark matter
has been accreted, i.e. approximately when $3 f_\text{\tiny PBH} (1+z/1000)^{-1} = 1$, or when tidal effects of nearby PBHs and other halos are efficient in destroying it.

Such a halo is characterised by a 
typical spherical density profile $\rho 
\propto r^{-\alpha}$ (with $\alpha \simeq 9/4$~\cite{Mack:2006gz, Adamek:2019gns} as confirmed by N-body simulations), truncated at a 
radius 
\begin{equation}
 r_h \simeq 0.019 \, {\rm pc} \left( \frac{M}{M_\odot} \right)^{1/3} \left( \frac{1+z}{1000}\right)^{-1}\,.
 \end{equation}
While direct accretion of dark matter onto the PBH is negligible~\cite{Ricotti:2007jk,Rice:2017avg}, the halo acts as a catalyst enhancing the baryonic accretion rate.

The presence of a dark halo clothing is usually taken into account in the accretion parameter  $\lambda$~\cite{Ricotti:2007jk}.
To estimate the hierarchy of scales in the problem, one can define the parameter
\begin{equation}
\kappa \equiv \frac{r_\text{\tiny B}}{r_h} = 0.22\left( \frac{1+z}{1000}\right) \left( \frac{M_h}{M_\odot}\right)^{2/3} \left( 
\frac{v_\text{\tiny eff}}{{\rm km \, s^{-1}}} \right)^{-2}\,.
\end{equation}
When $\kappa\geq2$, i.e. when the typical size of the halo is smaller than the Bondi radius, 
the accretion rate is the same as the one for a PBH of point mass $M_h$. On the other hand, if $\kappa <2$, only a fraction of the dark halo would be relevant for accretion, and 
 one has  to correct the quantities entering in the parameter $\lambda$ with respect to the naked case as~\cite{Ricotti:2007jk}
\begin{equation}
\beta^{h}_c \equiv \kappa^{\frac{p}{1-p}} \beta_c, \quad \lambda^{h} \equiv \bar\Upsilon^{\frac{p}{1-p}} 
\lambda 
(\beta^{h}_c), \quad \bar\Upsilon = \left( 1 + 10 \beta^h_c \right)^{\frac{1}{10}} {\rm exp} (2 - \kappa) \left( \frac{\kappa}{2} \right)^2\,,
\end{equation}
where $p = 2- \alpha \simeq -1/4$.
As shown in Refs.~\cite{DeLuca:2020bjf,Serpico:2020ehh}, a dark halo dress could enhance the accretion rate of few orders of magnitude, providing an important effect to be considered in the modelling of accretion.

\subsubsection{Beyond spherical symmetry}
\label{Sec1diskformation}

The previous results were based on the assumption of a spherically symmetric accretion flow onto the accreting PBHs.
Let us now discuss the accretion model when one goes beyond this assumption. While no complete theory of disk accretion exists from first principles, the standard scenario for disk formation requires that the accreted baryonic particles carry an angular momentum 
larger than the one at the innermost stable circular orbit. To understand the validity of this requirement, one can start from the expression of the baryon velocity variance $\sigma_{\text{\tiny b}} = \langle v_\text{\tiny b} \rangle$ provided in Refs.~\cite{1990ApJ...348..378O, Ricotti:2007au} as
\begin{equation}
\sigma_{\text{\tiny b}} \simeq (3.8 \times 10^{-7} {\rm km/s}) \, \xi^{-1.7}(z) \left( \frac{1+z}{1000} \right)^{-1} \left( \frac{M_h}{M_\odot} \right)^{0.85}\,,
\end{equation}
where  $\xi(z)= {\rm Max}[1, \langle v_\text{\tiny eff} \rangle/c_s]$ describes the effect of the PBH proper motion in reducing the Bondi radius. The above criterion can be recast in a condition for the gas velocity to be larger than the Keplerian velocity close to the PBH, $\sigma_{\text{\tiny b}} \gtrsim 2 D \xi^2 (z) c_s^2$,
in terms of a constant $D \sim \mathcal{O}(1 \div 10)$ describing relativistic corrections.
Using the expression for the dark halo mass shown in Eq.~\eqref{halo mass}, one can estimate the minimum PBH mass for which the accreting gas 
acquires a disk geometry~\cite{Ricotti:2007au, DeLuca:2020bjf},
\begin{equation}
\label{critM}
M \gtrsim 6\times 10^2 M_\odot \,D^{1.17} \xi^{4.33}(z) \frac{\left( 1+z/1000\right)^{3.35}}{\left[ 1 + 0.031 
\left(1+z/1000\right)^{-1.72} \right]^{0.68}}\,.
\end{equation}
It is interesting to also point out that, even without any velocity dispersion, the disk formation criterion can be satisfied if the nonlinear PBH motions at small scales are taken into account~\cite{Poulin:2017bwe}. In that case, the presence of near PBH neighbors may induce the formation of binaries, with size larger than the Bondi radius, which allow for a bulk motion of the baryonic gas with respect to the PBH pair center of mass. The corresponding condition for disk formation then reads~\cite{Poulin:2017bwe}
\begin{equation}
\sqrt{f_\text{\tiny PBH}} \left( \frac{M}{M_\odot} \right) \gg \left(\frac{1+z}{730} \right)^{3}\,,
\end{equation}
which shows that, whenever heavy PBHs constitute a large fraction of the dark matter, disk formation may happen at large redshifts because of their nonlinear motions.
Finally, notice that, contrarily to baryonic particles, the angular momentum of accreting dark matter is typically small and thus does not 
lead to the formation of a disk, while it impacts on the density profile of the dark halo which envelops the PBH.

Besides the necessary condition derived in Eq.~\eqref{critM}, the disk geometry also depends on the strength of the accretion rate, $\dot{m}$. 
Different regimes of $\dot{m}$ can be considered. First, if accretion is not efficient enough, $\dot m<1$, but the geometry is non-spherical (when Eq.~\eqref{critM} is satisfied), an advection-dominated accretion flow~(ADAF) may form~\cite{Narayan:1994is}. On the other hand, when $\dot 
m\gtrsim1$, non-spherical accretion can give rise to a geometrically thin accretion disk~\cite{Shakura:1972te}. Finally, for an extremely large accretion rate, $\dot m\gg1$, the accretion luminosity might be strong enough that the disk ``puffs up'' and becomes thicker. These regimes will be discussed in more details in the next Section.

For the PBH masses we are interested in, $\dot m$ never exceeds unit significantly (see discussion in Section~\ref{Sec3}), such that the thin-disk approximation should be reliable in the slightly super-Eddington regime of PBHs. Notice also that the condition $\dot m\gtrsim1$ is always more stringent than 
the requirement~\eqref{critM}, and it can therefore be considered as a sufficient condition for the formation of a thin disk around an isolated PBH~\cite{DeLuca:2020bjf}.

\subsubsection{Bolometric luminosity}
Accreting PBHs are responsible of injecting radiation into the surrounding medium. One can estimate the amount of emitted energy by computing the PBH bolometric luminosity, which is a necessary ingredient to study the role of the emitted radiation in the heating and ionization of the intergalactic medium.

The bolometric luminosity of an accreting PBH depends on the matter accreted onto the PBH per unit time $\dot{M}$, as well as on an overall efficiency factor $\epsilon$ related to the conversion of accreted matter into radiation. It reads
\begin{equation}
L = \epsilon \dot{M} = \epsilon \dot{m} L_\text{\tiny Edd}\,,
\end{equation}
in terms of the Eddington luminosity $L_\text{\tiny Edd}$, that provides an upper bound at which spherical accretion is balanced by radiation pressure, and defined as
\begin{equation}
L_\text{\tiny Edd} = \frac{4 \pi \mu G M m_p}{\sigma_\text{\tiny T}} \simeq 1.3 \cdot 10^{38} \left( \frac{M}{M_\odot}\right) {\rm erg/s}\,.
\end{equation}
The efficiency factor $\epsilon$ depends on the accretion rate $\dot{m}$ and on the geometry of the accretion flow, so that $\epsilon = \epsilon (\dot{m})$. Its typical value is of the order of $\mathcal{O}(0.1)$ for spherical accretion and, in order to satisfy the Eddington bound, decreases as $\dot M$ increases.

Solving for the behaviour of $\epsilon (\dot{m})$ is in general a complicated radiative transfer problem. However, following the description discussed above, an approximate understanding can be outlined for different geometries of the accretion flow~\cite{Ricotti:2007au}.
For spherical accretion, the dominant
emission mechanism is bremsstrahlung, with most of the radiation originating from the region just outside the event horizon.
In this case, the efficiency to convert rest-mass energy into
radiation is proportional to $\dot{m}$ and it is given by $\epsilon=0.011 {\dot m}$ \cite{1973ApJ...180..531S,1973ApJ...185...69S} (even smaller values can be achieved when Compton cooling is negligible~\cite{Ali-Haimoud:2016mbv}).

When the gas angular momentum is non negligible (i.e. when the condition \eqref{critM} is satisfied), different regimes arise depending on the gas cooling efficiency and accretion rate $\dot m$. For example, when $\dot m < 1$, the accretion flow is typically not dense enough for electrons to cool efficiently, such that most of the rest-mass energy is not radiated away and is advected into the central BH, giving rise to an advection-dominated accretion flow (ADAF)~\cite{Ichimaru:1977uf, Yuan:2014gma}. In this case the accretion disk is hot, geometrically thick and optically thin, and the radiative
efficiency is linearly suppressed and may reach values $\epsilon \simeq 0.1 \dot m$~\cite{Narayan:1994is}. 
On the other hand, if the radiative cooling of the gas is efficient and $\dot m \gtrsim 1$, theoretical models suggest the formation of thin-disk geometries~\cite{Shakura:1972te}. In this case, the maximal energy per unit mass available is uniquely determined by the binding energy at the innermost stable orbit, corresponding to a
radiative efficiency approaching a constant value $\epsilon \simeq 0.1$~\cite{Ricotti:2007au}. In this case therefore the bolometric luminosity can never
exceed the Eddington limit.
Finally, for larger accretion rates $\dot{m} > 1$, several phenomena may come into play. For instance, radiative and kinetic feedback can break stationary conditions, generating outflows or periods of high luminosity alternating with periods of low luminosity. Otherwise, phases with quasi-steady state, super-Eddington, mass accretion can take 
place, with a corresponding drop in efficiency to satisfy $L \lesssim L_\text{\tiny Edd}$.
The impact of feedback effects on the accretion rate is therefore crucial to understand the effective geometry of disks around PBHs; in particular, if feedback phenomena are efficient enough in suppressing the rate below the Eddington limit, then one would expect accreting PBHs
 during the dark ages to form an ADAF~\cite{Poulin:2017bwe, Agius:2024ecw}. While a complete understanding of their effects is still under investigation, we provide a more detailed discussion in the next Section and in Chapter 26.

The efficient emission of radiation in the surrounding medium could modify its properties, resulting eventually into its heating and ionization, and breaking the assumption of constant ionization fraction over the accretion flow. In particular, close enough to the PBH, the gas may eventually become fully ionized through photoionizations by the outgoing radiation field~\cite{Ali-Haimoud:2016mbv}.  By solving the time evolution of the ionization fraction coupled to the one for the gas temperature, one could get a more precise estimate of the gas accretion rate onto PBHs and the corresponding bolometric luminosity. The interested reader can find dedicated discussions on these effects in Refs.~\cite{Ali-Haimoud:2016mbv,Zhang:2023hyn,Ziparo:2022fnc} and Chapter 26.

\subsection{Accretion onto PBH binaries}
\label{Sec1binaries}

After their formation in the early universe, PBHs are dynamically coupled to the cosmic expansion, with negligible peculiar velocities. When their local density becomes comparable to the one of the surrounding radiation fluid, usually after the matter-radiation equality, they can become gravitationally bound to each other. However, because of the presence of large Poisson fluctuations at small scales, some PBHs can decouple even earlier. When they decouple, the closest PBH pairs would start to fall towards each other in a head-on collision, unless the torque caused by the gravitational field of the surrounding PBHs and other matter inhomogeneities prevents it. In that case the pair would give rise to a binary system~\cite{Nakamura:1997sm, Ioka:1998nz, Raidal:2018bbj}.

From that epoch on, PBH binaries could experience a phase of baryonic accretion, as it happens for individual PBHs. In this case, however, one has to take into account both local processes, onto the individual PBHs in the binary, and global processes, i.e. on the binary as a whole. 
In this Section we summarise how accretion proceeds for PBH binaries following Refs.~\cite{DeLuca:2020bjf,DeLuca:2020qqa}.

\begin{figure}[t]
\centering
\includegraphics[scale=.3]{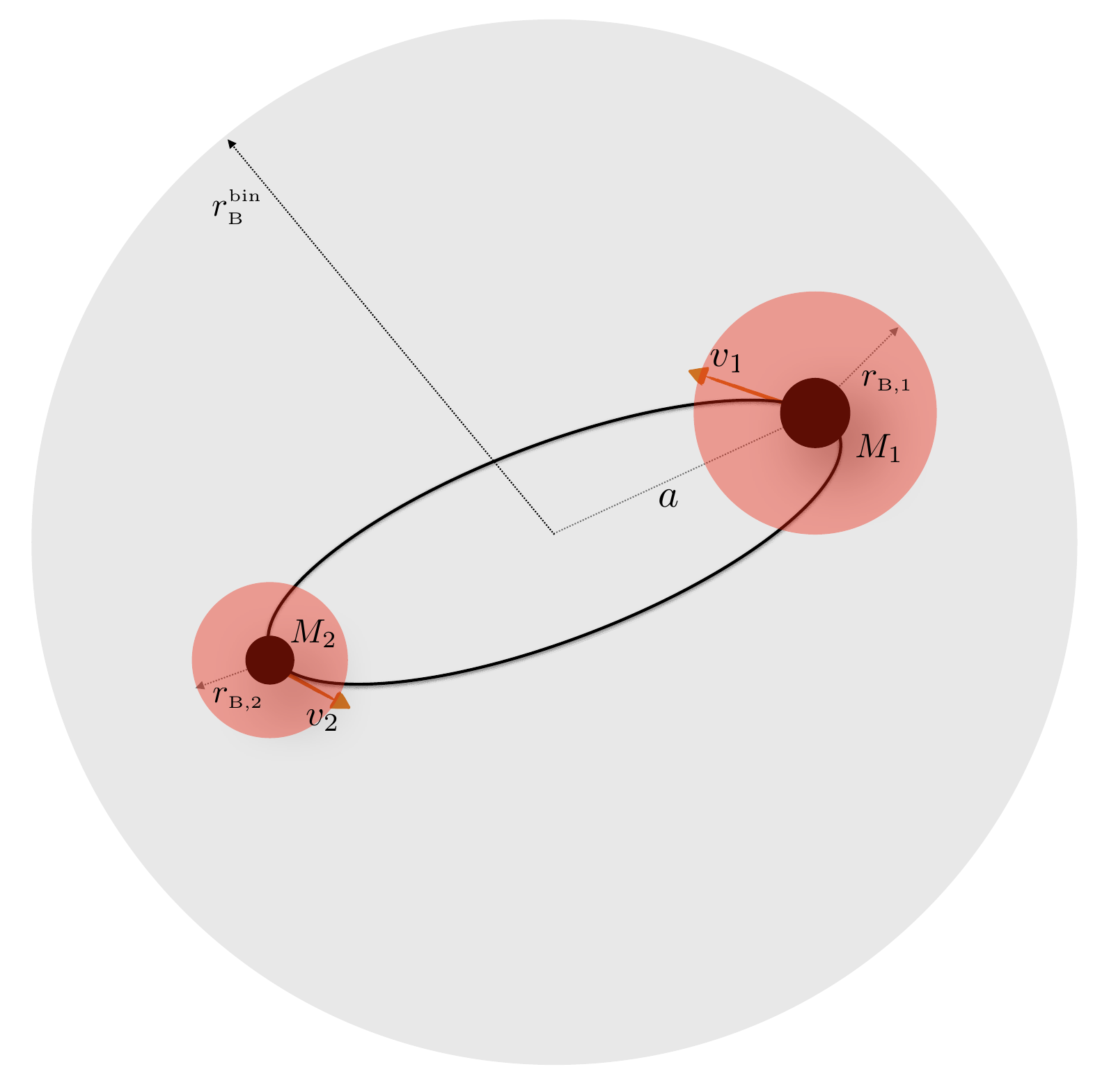}
\caption{Pictorial representation of a PBH binary undergoing a phase of baryonic accretion. The different scales of the problem are shown, including the Bondi radius of the binary which is taken to be much bigger than the orbital separation.
Figure adapted from Ref.~\cite{DeLuca:2020qqa}.}
\label{fig:draw}       
\end{figure}

Let us consider a PBH binary with total mass $M_\text{\tiny tot} = M_1 + M_2$, mass ratio $q=M_2/M_1\leq1$, reduced mass $\mu = M_1 M_2 /(M_1 + M_2)$, and orbital parameters $a$ and $e$, describing the semi-major axis and eccentricity, respectively. The setup is schematically depicted in Fig.~\ref{fig:draw}.

Similarly to the case of individual PBHs, one can identify a Bondi radius of the binary
\begin{equation}
	r^\text{\tiny bin}_\text{\tiny B} = \frac{G M_\text{\tiny tot}}{v^2_\text{\tiny eff}}\,, 
\label{rBbin}
\end{equation}
in terms of the effective velocity $v_\text{\tiny eff}= \sqrt{c_s^2 + v^2_\text{\tiny rel}}$,  with which the binary center of mass moves relative to the surrounding gas.
The presence of another scale, the binary semi-major axis $a$,  identifies two different regimes of interest for the accretion process: 
for large semi-major axis, $a\gg r^\text{\tiny bin}_\text{\tiny B}$, accretion occurs onto the single PBHs independently (each 
moving at a characteristic velocity that depends on the orbital one). In this case the description outlined in the previous Section would apply. On the other hand, as
the binary hardens and the orbital separation decreases, the binary Bondi radius would become the dominant scale ($a\ll
r^\text{\tiny bin}_\text{\tiny B}$) and change the properties of accretion. We will discuss this case below. Notice that this could occur much before  gravitational waves emission becomes the dominant mechanism for driving the inspiral evolution.

The corresponding accretion rate for a PBH binary reads
\begin{equation}
 \label{R1bin}
\dot M_\text{\tiny bin} = 4 \pi \lambda \rho_\text{\tiny gas} v_\text{\tiny eff} r^{\text{\tiny bin}\, 2}_\text{\tiny B}\,.
\end{equation}
Using the geometrical properties of the binary system, we can now derive how accretion proceeds for each binary component.
The PBH positions and velocities with respect to the center of mass are given by~\cite{poisson_will_2014}
\begin{equation}
r_1 = \frac{q}{1+q}r, \qquad v_1 = \frac{q}{1+q}v;  \qquad r_2 = \frac{1}{1+q}r, \qquad v_2 = \frac{1}{1+q}v  
\end{equation}
as a function of their relative distance and velocity
\begin{equation}
r = a (1-e \, {\rm cos}u), \qquad v = \sqrt{G M_\text{\tiny tot} \left( \frac{2}{r}-\frac{1}{a}\right)}\,,
\end{equation}
and orbital angle $u$. The latter evolves in time following the law $\sqrt{a^3/G M_\text{\tiny tot}}(u(t)-e \sin u(t))=t-C$, 
in terms of an integration constant  $C$~\cite{poisson_will_2014}. The corresponding effective velocities for each PBH component are given by
\begin{equation}
v_\text{\tiny eff,1} = \sqrt{v^2_\text{\tiny eff} + v^2_1 }\,, \qquad  v_\text{\tiny eff,2} = \sqrt{v^2_\text{\tiny 
		eff} + v^2_2}\,. 
\label{veffi}
\end{equation}
Since we are considering a situation where the typical binary semi-axis is much smaller than the Bondi radius of the binary (as shown in 
Fig.~\ref{fig:draw}), the total infalling flow of baryons towards the binary is constant, i.e.
\begin{equation}
4 \pi \rho_\text{\tiny gas}(R) v_\text{\tiny ff}(R)R^2 = {\rm const} = \dot M_\text{\tiny bin}\,,
\end{equation}
where the gas free fall velocity, $v_\text{\tiny ff}$, is computed by assuming that at large distances from the binary, $R \sim 
r^\text{\tiny bin}_\text{\tiny B} \gg a$, it reduces to the usual effective velocity $v_\text{\tiny eff}$, i.e.~\cite{DeLuca:2020qqa}
\begin{equation}
v_\text{\tiny ff} (R) = \sqrt{v^2_\text{\tiny eff} + \frac{2G M_\text{\tiny tot}}{R} - \frac{2G M_\text{\tiny 
			tot}}{r^\text{\tiny bin}_\text{\tiny B}}}\,.
\end{equation}
The gas density profile $\rho_\text{\tiny gas}(R)$ at a distance $R$ from the binary center of mass then reads
\begin{equation}
\rho_\text{\tiny gas}(R) = \frac{\dot M_\text{\tiny bin}}{4\pi v_\text{\tiny ff}(R)R^2 }\,,
\end{equation}
showing that a constant flow of infalling baryons towards the binary results into an increase in the density compared to 
 its mean cosmic value at the binary Bondi radius, $\rho_\text{\tiny gas}$.

From the definition of the gas density profile $\rho_\text{\tiny gas} (R) $, one can then write down the accretion rates for the individual binary components in terms of their effective Bondi 
radii $r_\text{\tiny B,i} = G M_i/ v_\text{\tiny eff,i}^2$ as
\begin{equation}
\dot M_1 = 4 \pi G^2  \rho_\text{\tiny gas}(r_\text{\tiny B,1}) v^{-3}_\text{\tiny eff,1} M_1^2\,, \qquad 
\dot M_2 = 4 \pi G^2  \rho_\text{\tiny gas}(r_\text{\tiny B,2}) v^{-3}_\text{\tiny eff,2} M_2^2\,. 
\label{Rindiv}
\end{equation}
Few considerations are in order. First, one can notice that, since the local PBH velocities shown in Eq.~\eqref{veffi} are of the order of the 
	orbital velocity $v$, which is much larger than $v_\text{\tiny rel}$ and $c_s$ when $a \ll r^\text{\tiny bin}_\text{\tiny B}$, the individual Bondi radii $r_\text{\tiny B,i}$ are much smaller than the binary Bondi radius $r^\text{\tiny bin}_\text{\tiny B}$. Second, in the above expressions we have assumed the individual PBHs to be naked, fixing their accretion  parameter $\lambda \approx 1$, 
while we have described the whole binary as clothed by a dark halo, with parameter $\lambda$ that takes 
into account the ratio 
of the binary Bondi radius compared to the dark halo radius~\cite{Ricotti:2007au}, $r^\text{\tiny bin}_\text{\tiny B}/r^\text{\tiny bin}_h$, as discussed in the previous Section.
The assumption of neglecting the dark halos for single PBHs is based on the consideration that, during the orbital evolution, the halos surrounding each PBHs would get destroyed by tidal effects, leaving one single halo surrounding the binary~\cite{Kavanagh:2018ggo, Coogan:2021uqv}.

After some simplifications, the individual PBH accretion rates read
\begin{align}
	& \dot M_1 = \dot M_\text{\tiny bin} \sqrt{\frac{1+\zeta+(1-\zeta) \gamma ^2}{2 (1+\zeta) 
			(1+q)+(1-\zeta)(1+2 q) \gamma ^2}}\label{M1dotGEN}\,, \\
	& \dot M_2 =  \dot M_\text{\tiny bin} \sqrt{\frac{(1+\zeta) q+(1-\zeta) q^3 \gamma ^2}{2 (1+\zeta)
			(1+q)+(1-\zeta) q^2 (2+q) \gamma ^2} }\,, \label{M2dotGEN}
\end{align}
in terms of $\zeta=e\cos u$ and $\gamma^2=a v^2_\text{\tiny eff}/ \mu q$. Since the orbital period is usually much smaller than the accretion timescale $\tau_\text{\tiny acc}$, one can average over the angle $u$, eliminating the time dependence, and also notice that the dependence on the eccentricity is negligible. 
Finally, since $G M_1 q^2/a \gg v^2_\text{\tiny eff}$ for PBHs heavier than a solar mass and for large enough semi-major axis, one can further simplify the expressions to~\cite{DeLuca:2020qqa}
\begin{align}
	\dot M_1 = \dot M_\text{\tiny bin}  \frac{1}{\sqrt{2 (1+q)}}\,, \qquad \dot M_2 = \dot M_\text{\tiny bin}  \frac{\sqrt{q} }{\sqrt{2 (1+q)}}\,,
\label{M1M2dotFIN}
\end{align}
which recover the expected behaviour $\dot M_1 = \dot M_2 =  \dot M_\text{\tiny bin}/2$ in the limit of equal PBH masses $q \to 1$.
These equations therefore provide the accretion rate for the PBH binary components. Reintroducing the Eddington normalised rates, $\dot m_i = \tau_\text{\tiny Salp} \dot M_i/M_i$, one gets 
\begin{align}
	\dot m_1 =  \dot m_\text{\tiny bin} \sqrt{\frac{1+q}{2}}\,, \qquad \dot m_2 =  \dot m_\text{\tiny bin} 
	\sqrt{\frac{1+q}{2q}}\,. \label{m1m2dotFIN}
\end{align}
From these equations, one can also derive the time evolution for the binary mass ratio, which reads~\cite{DeLuca:2020qqa}
\begin{equation}
\label{massratioevo}
\dot q = q \left( \frac{\dot M_2}{M_2} -  \frac{\dot M_1}{M_1} \right) = \frac{q}{\tau_\text{\tiny Salp}} \left( \dot m_2 - \dot 
m_1 \right)\,.
\end{equation}
Since $\dot m_2>\dot m_1$ as shown in Eq.~\eqref{m1m2dotFIN}, this equation proves that the mass ratio grows until it reaches a stationary point when $q=1$ and $\dot 
m_1=\dot m_2$. In other words, accretion onto a PBH binary pushes the binary masses to balance each other on secular time scales.

Let us close this Section by discussing the formation of an accretion disk for PBHs in binary systems. As shown above, each binary component has a velocity of the order of the orbital velocity $v_{1,2} \sim v$, which is much larger than the binary effective velocity $v_\text{\tiny eff}$. Although the Bondi radius of the individual PBHs is much smaller than the binary Bondi radius, it is still parametrically larger than the ISCO radius, $r_\text{\tiny B,i} > r_\text{\tiny ISCO,i}$, such that conservation of angular momentum along the accretion flow implies that the condition for the formation of a disk is always satisfied (see Eq.~\eqref{critM})~\cite{DeLuca:2020qqa}. Furthermore, the condition $\dot m \gtrsim 1$, to ensure the formation of a thin accretion disk, is more easily fulfilled by PBHs in binaries rather than by individual ones, since accretion for binaries is enhanced by the larger total mass of the system. 
In Section~\ref{Sec3} we will show the effects of baryonic accretion on the mass and spin parameters of PBHs which are either isolated or in binaries.

\section{Feedback effects}
\label{Sec2}

So far we have considered the dynamics of an accreting PBH without considering the effect of emission on the surrounding environment.
However, accurate modelling of accretion requires considering the reciprocal interaction between the PBH and the surrounding medium.
Feedback effects describe this intertwined relation between the PBH accretion and emission mechanism, and the gas medium.
The importance of feedback effects is well understood in other branches of Astrophysics, for instance in galaxy dynamics, where AGN feedback turns out to be a key ingredient in explaining galaxy evolution, see, e.g., Ref.~\cite{2017FrASS...4...42M} for a review on the subject.
In this Section we will discuss the different type of feedback effects which may arise during a PBH accretion phase.

\subsection{Local thermal and ionization feedback}

Local feedback effects refer to a set of physical effects the PBH has on its local environment, where ``local'' in this context refers to the PBH sphere of influence.
The solution of equations~\eqref{eq:mass}, \eqref{eq:momentum} and~\eqref{eq:heat-compt} describes the temperature profile of the gas around the PBH, and in particular it describes the temperature  at the Schwarzschild radius depending on the gas ionization scheme assumed~\cite{Ali-Haimoud:2016mbv}.
Two common ionization schemes are typically assumed, ``collisional ionization'' and ``photoionization''.

In the former scenario, the temperature of the gas grows due to the adiabatic infall towards the PBH until it reaches the ionization temperature~$T_\mathrm{ion}\approx 10^4\ \mathrm{K}$.
At that stage, all the energy gained by the infall dynamics is converted into ionizing the gas, and only when the gas becomes fully ionized the temperature starts growing again.
On the other hand, in the latter scenario, ionization is driven by high-energy photons emitted by ultra-relativistic electrons close to the Schwarzschild radius, thus the growth of the temperature does not need to stop while the gas is getting ionized, resulting in a higher temperature~$T_S$ of the gas close to the compact object.
In both simplified pictures the temperature grows as~$T(r)\propto r^{-n}$, where the adiabatic index is~$n\approx 1$ ($n \approx 2/3$) when $T \lesssim m_e$ ($T \gtrsim m_e$).
The photoionization scenario is characterized by a higher temperature, around~$T_S \approx 10^{11}\ \mathrm{K}$, than the one in the collisional ionization scenario ($T_S \approx 10^9\ \mathrm{K}$), thus in a higher luminosity of the PBH.

In the simplified picture described above it was assumed that heating of the infalling gas is only due to compression. 
On the other hand, the radiation produced by the PBH can heat the surrounding material also through Compton scattering, inducing a local thermal feedback effect. 
It turns out that such Compton heating is comparable to the standard adiabatic heating, or to the Compton cooling to CMB photons at scales comparable to the Bondi radius~\cite{Ali-Haimoud:2016mbv}, and it is therefore negligible. 
Similarly, whether the plasma around the PBH is optically thin or thick also has an impact in the transmission of heating across the surrounding environment.
However, it can be showed that as long as the accretion rate is smaller than the Eddington limit, i.e., $\dot{m}\ll 1$, the plasma is optically thin and the previous modelling is not affected by this assumption~\cite{Ali-Haimoud:2016mbv}.

Finally, let us stress that also the choice of either of the two simplifying ionization schemes cannot fully describe the realistic case. This can be easily checked by analysing the size of the Str\"omgren sphere around an accreting PBH~\cite{Ali-Haimoud:2016mbv}.
In particular, the collisional ionization picture predicts a radius of such sphere that is smaller than the prediction obtained from detailed balance argument~\cite{2006agna.book.....O}, while in the case of photoionization the radius is found to be larger.
Thus, the realistic scenario should fall between these two possibilities, and it is heavinly impacted the feedback provided by the accreting PBH.

\subsection{Global feedback}

Due to the supersonic motion of DM across the cosmic medium at high redshifts~\cite{2010PhRvD..82h3520T}, the geometry of accretion cannot be purely spherical, even at scales of the order of the Bondi radius.
Instead, an axisymmetric accretion column develops behind the PBH, through which the medium is channeled directly to the accretor.
Despite this fundamental difference in the geometry, accretion still proceeds at a rate comparable to the Bondi-Hoyle-Littleton one~\cite{1944MNRAS.104..273B, hoy39} described in Section~\ref{Sec1}.

Conversely, in front of the compact object, we observe the formation of a bow-shaped ionization front on scales larger than the Bondi radius for  PBH masses~$M_\text{\tiny PBH} \gtrsim 10^2 M_\odot$~\cite{Ricotti:2007au, Park_2013, 10.1093/mnras/staa1394}, which distinguish the neutral region ahead of the front from the region behind it, which embeds the PBH, where the medium has been ionized.
Due to the presence of these fronts, the gas density close to the PBH is reduced, inducing a lower accretion rate than in the pure spherical case~\cite{Facchinetti:2022kbg}. As shown in Ref.~\cite{Park:2012cr}, this conclusion seems to be valid also for relatively slow PBHs, and even in the presence of dark matter halos or accretion disks~\cite{Agius:2024ecw}, strongly impacting the efficiency and phenomenology of the accretion process.
Moreover, gas inside these ionization bubbles will get heated, and as soon as its temperature reaches values higher than the background one, accretion is further reduced, especially at redshifts lower than recombination.

On the other hand, the stability of this kind of structure is non-trivial, since they can be disrupted due to the high speed motion of the compact object~\cite{10.1093/mnras/staa1394}.
In that case the PBH can experience bursts in accretion due to the sudden inflow of cold gas, that can be easily accreted, even though the dynamics of this process is highly chaotic.

\subsection{Mechanical and radiative feedback}

Because of their interaction with the surrounding environment, PBHs can generate complex structures called outflows, winds and/or jets, depending on their level of collimation, that extend from the region close to the event horizon up to scales many times larger than the Bondi radius, see, e.g., the seminal Refs.~\cite{bla77, bla82}.
These outflows may modify the accretion efficiency both through mechanical or radiative processes, which effectively sweep the medium away from the compact object, or simply just heat it.
While the role of radiative feedback is harder to model because it strongly depends on the details of the outflow and the medium, mechanical feedback processes are more robust to small changes in the details of the modelling, and they have already been studied in multiple contexts, ranging from stars, stellar mass BHs, supermassive BHs and galaxies, see, e.g., Refs.~\cite{2016NewAR..75....1S, 2017MNRAS.470.3332I, 2018MNRAS.473.2673L, 2019MNRAS.482.4642Z, 2020MNRAS.492.2755G, 2020MNRAS.494.2327L}.

On the other hand, the role of mechanical feedback on PBH accretion has been explored only recently~\cite{Bosch-Ramon:2020pcz, 2022A&A...660A...5B, Piga:2022ysp}.
In this scenario, the presence of a magnetic field and/or some excess of thermal energy in the accreted gas in the proximity of the compact object can launch supersonic outflows with different degrees of collimation.
Even under the conservative assumption that these outflows carry only a small fraction of the accreted energy, such structures can escape the vicinity of the PBH, overcome the ram pressure of the inflowing gas, reach scales larger than the Bondi radius and sweep away and/or heat the gas.
The level of effectiveness of this process depends significantly on the angle between the PBH direction of motion and the outflow direction, in particular it is maximised if the two are aligned.
This pure mechanical effect reduces accretion because the hot medium needs time to cool down and approach again the PBH.

The dynamics of this process is highly chaotic, and numerical simulations are required to derive the exact amount of accreted gas~\cite{2020MNRAS.492.2755G, 2020MNRAS.494.2327L, Bosch-Ramon:2020pcz, 2022A&A...660A...5B}.
Nevertheless, under conservative assumptions, and for choices of gas density and PBH velocity that resemble the conditions present during the epochs after the cosmological recombination era and previous to structure formation, it was showed that mechanical feedback alone can reduce the accretion rate by an order of magnitude.
Finally, the outflow itself could irradiate energy that would eventually be deposited into the surrounding medium; however, the implications of this feedback effect and its role on the accretion dynamics have been studied only in the context of PBH accreting in virialized structures, see, e.g., Ref.~\cite{Takhistov:2021upb}.

\subsection{Feedback mechanisms in the late Universe}

In summary, modelling in a self-consistent manner the accretion and subsequent emission of radiation is one of the open questions in astrophysics.
This Chapter covered the basics of feedback mechanisms that have mostly been explored in the high-redshift Universe, previously to the formation of virialized structures.

At the time of their formation, a part of the PBH population would start falling in their gravitational potential wells around redshifts $z \simeq \mathcal{O}(10)$, increasing the relative velocity up to one order of magnitude~\cite{Hasinger:2020ptw}. This will result in a consequent suppression of the accretion rate.

Other examples of feedback have been considered in the literature, for instance concerning the gravitational role played by the gas surrounding the PBH~\cite{Ricotti:2007jk, Serpico:2020ehh, Takhistov:2021upb}, the possibility of accreting at super-Eddington rates~\cite{Takeo:2017qbj, Takeo:2019uef} (where the formation of thick disks may impact on the accretion luminosity and feedback), or the possibility of having PBH not evolving ``in isolation''~\cite{Ali-Haimoud:2016mbv}.
However, all these feedback mechanisms are effective only for PBHs with masses~$M_\text{\tiny PBH} \gtrsim 10^4\ M_\odot$, where the picture presented in Section~\ref{Sec1} cannot be used to  accurately describe them. 

We conclude by noting that once virialized structures form, the study of PBHs in non-linear environments becomes even more challenging, and can be reasonably done only with the help of complex and model-dependent numerical simulations, without relying on simple analytic prescriptions.
In this sense we parameterise the uncertainties in certain aspects of the accretion model by introducing an effective cut-off redshift~$z_\text{\tiny cut-off}$, after which PBH accretion is not efficient anymore, and limiting its effectiveness to the pre-Cosmic Dawn era~\cite{DeLuca:2020qqa}.
The exact value of this cut-off redshift is treated as a phenomenological parameter in Section~\ref{Sec3}.

\section{Effects of accretion on a PBH population}
\label{Sec3}
Gas accretion provides the main effect driving the mass and spin evolution of PBHs during their cosmic evolution. In particular, at redshifts $z \in (10 \div 100)$, PBHs of certain mass may experience an efficient accretion phase, provided that they are surrounded by a dark matter halo which catalyse the process, and that feedback effects are not strong enough in suppressing the rate. From the discussion in the previous Section, we will implement $z_\text{\tiny cut-off}$ as a parameter describing the duration and efficiency of the accretion phase. For low enough cut-off redshifts, one can show that the late time PBH mass and spin distributions might be significantly different from those at high redshifts. 
In the following we summarise the main effects of accretion onto a PBH population. The interested reader can find further details in Refs.~\cite{DeLuca:2020bjf,DeLuca:2020qqa, Franciolini:2021xbq}.

\subsection{PBH mass evolution}

For PBHs in the solar mass range, baryonic accretion is mostly ineffective at redshifts $z\gtrsim 100$, because of the longer accretion timescale compared to the 
age of the universe at that epoch. On the other hand, at smaller redshifts, PBHs may start evolving rapidly if they are heavy enough, $M\gtrsim {\cal O}(10) M_\odot$.
A PBH, formed at  redshift $z_\text{\tiny i}$ with initial mass $M^{_\text{\tiny i}} \equiv M (z_\text{\tiny i})$, may reach a final mass $M (M^\text{\tiny i}, z_\text{\tiny cut-off})$, obtained by solving the equation
\begin{equation}
\label{Mass evolution}
\frac{d M}{d t} = \dot{M}_\text{\tiny B} (M,t) = \dot{M}_\text{\tiny Edd} \dot{m}, \quad z (t) \in [z_\text{\tiny i} ; z_\text{\tiny cut-off}], \, M (t) \in [M^\text{\tiny i} ; M  (M^\text{\tiny i}, z)]\,,
\end{equation}
depending on the actual redshift $z_\text{\tiny cut-off}$ at which structure formation and reionization take place, as well as on feedback effects.

\begin{figure}[t!]
\centering
\includegraphics[scale=.37]{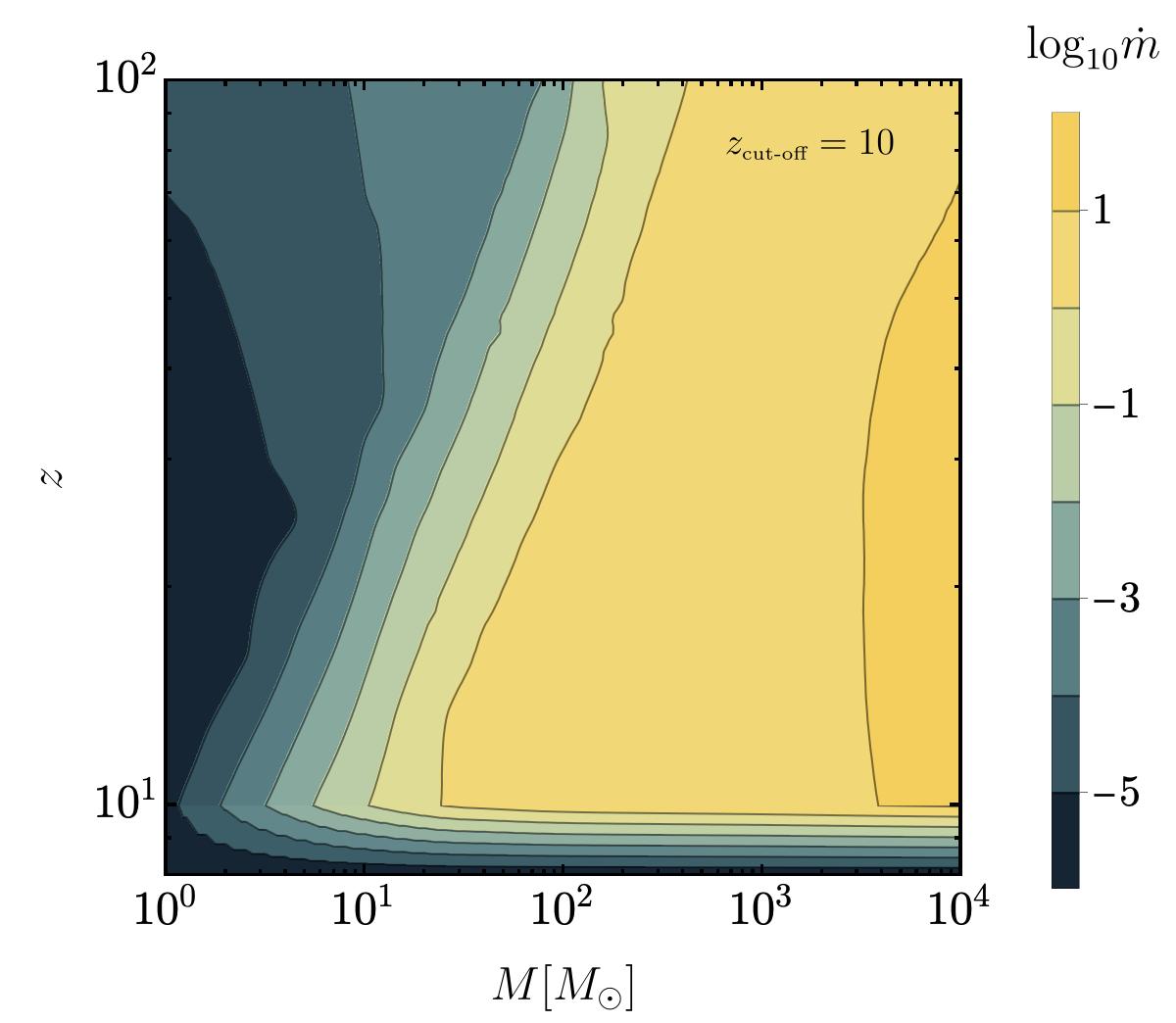}
\caption{Accretion rate $\dot m$ as a function of the PBH mass $M$ and redshift $z$, assuming a cut-off redshift $z_\text{\tiny cut-off} = 10$ after which the accretion efficiency drops. Figure adapted from Ref.~\cite{DeLuca:2020bjf}.}
\label{fig:mdot}       
\end{figure}
 
Following the Bondi formalism discussed in Section~\ref{Sec1}, an illustrative plot of the accretion rate $\dot{m}$ in terms of the PBH mass $M$ and redshift $z$, and assuming a sharp efficiency decrease after $ z_\text{\tiny cut-off} \sim 10$, is shown in Fig.~\ref{fig:mdot}. Based on the seminal work~\cite{Ricotti:2007au}, and including the effects of a dark matter halo as thoroughly discussed in Section~\ref{Sec1}, it shows that PBH accretion may be efficient in some regions of the parameter space, $\dot{m} \gtrsim 1$, for $M \gtrsim \mathcal{O}(30) M_\odot$, in the absence of a significant radiative feedback. For masses larger than about $10^4 M_\odot$, feedback effects are crucial in the modelling of the accretion rate, and more involved computations are needed to provide a definite behaviour. A different choice in the cut-off redshift amounts to a drastic decrease of $\dot{m}$ at that corresponding value.

\begin{figure}[t!]
\centering
\includegraphics[scale=.5]{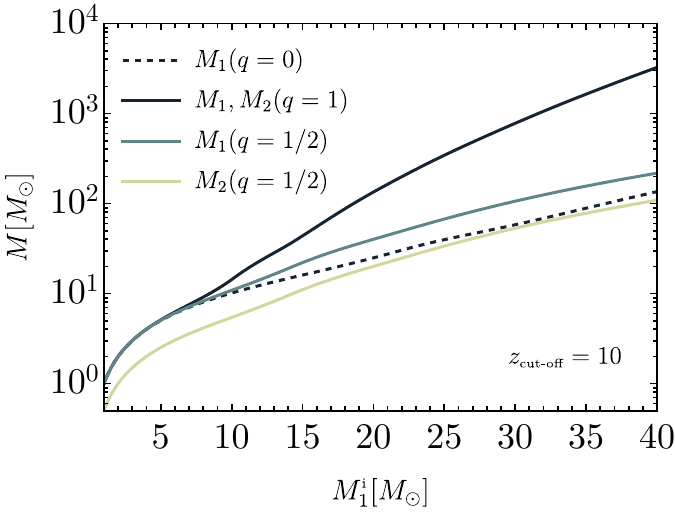}
\includegraphics[scale=.5]{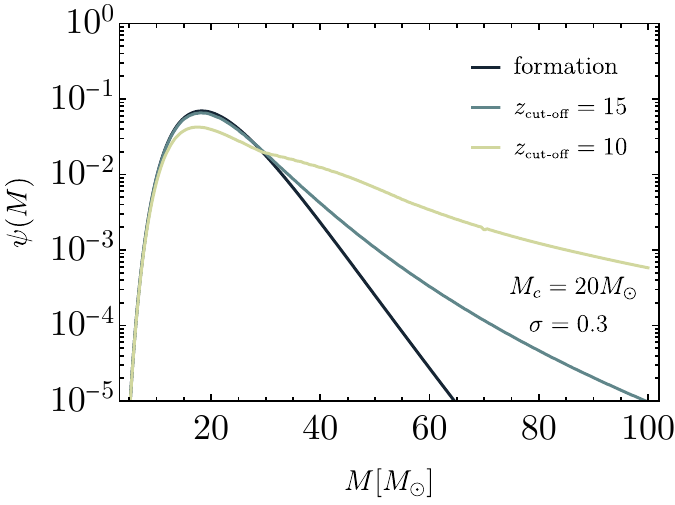}
\caption{Left panel: Behaviour of the PBH final masses $M$ in terms of the initial ones $M^\text{\tiny i}$, assuming different values for the final mass ratio $q$. 
We have fixed the cut-off redshift to the representative value $z_\text{\tiny cut-off} = 10$. The dashed black line shows the evolution for a single PBH, while the solid lines hold for PBH binaries.
Right panel: Evolution of the PBH mass function $\psi (M)$, assuming an initial lognormal shape with central mass $M_c = 20 M_\odot$ and variance $\sigma = 0.3$, for different values of the cut-off redshift $z_\text{\tiny cut-off} = 15,10$. 
Figure adapted from Ref.~\cite{DeLuca:2020qqa}.}
\label{fig:masses}       
\end{figure}

Once the accretion rate $\dot{m}$ for individual PBHs is known, and using the extended description for accretion when PBHs are assembled in binaries (see Sec.~\ref{Sec1binaries}), one can solve Eq.~\eqref{Mass evolution} to get the evolution of the PBH mass, which is shown in the left panel of Fig.~\ref{fig:masses}. In particular, since the initial binary parameters are unmeasurable, it is more relevant to analyse the dependence of the final masses $M$ at redshifts after the cut-off, for each value of the initial PBH mass $M^{\text{\tiny i}}$. The black dashed line shows the case of an isolated PBH, while the solid lines show the final masses $M_1,M_2$ of a PBH binary, varying the initial primary component $M^{\text{\tiny i}}_1$ and assuming different values for the final mass ratio, $q = M_2/M_1$. The black solid line shows the case of an equal mass binary $q =1$, while the blue and yellow lines show the case when $q = 1/2$.

Few comments follow. First, PBHs with initial masses larger than $\mathcal{O}(10) M_\odot$ experience an efficient growth, increasing their masses of few orders of magnitude at small redshifts, under the presence of accretion-boosting dark matter halos and neglecting radiative feedback. These objects could be natural candidates for intermediate-mass BHs. Second, PBHs in binaries are expected to grow more efficiently compared to individual PBHs, since the binary total mass is responsible for a larger accretion rate compared to single PBHs. Notice, in particular, the growth of initially nearly-equal mass binaries (solid black) compared to single PBHs (dashed black).
Finally, by comparing the blue and yellow lines, one can easily realise that the secondary component in the binary always experiences a relative stronger growth compared to the primary. This result gives the expectation that 
strongly-accreting binaries tend to have final mass ratios close to unity, see Eq.~\eqref{massratioevo}.

The evolution of PBHs under accretion affects as well  their mass distribution in the universe. 
The mass function $\psi(M,z)$ is usually defined as the fraction of PBHs with mass in the interval $(M, M + {\rm d} M)$ at redshift $z$. For an initial $\psi (M^\text{\tiny i},z_\text{\tiny i})$, its evolution is governed 
by~\cite{DeLuca:2020bjf,DeLuca:2020fpg}
\begin{equation}
	\psi(M(M^\text{\tiny i},z),z) {\rm d} M = \psi(M^\text{\tiny i},z_\text{\tiny i}) {\rm d} M^\text{\tiny i}\,.
 \label{psiev}
\end{equation}
The latter is mostly dictated by isolated PBHs, which dominate the total PBH abundance in the universe. The main effect of accretion on the mass distribution is to broaden it at high masses, giving rise to a high-mass tail that can be orders of magnitude above its formation value~\cite{DeLuca:2020fpg}. A representative example is shown in the right panel of Fig.~\ref{fig:masses}, where we have assumed an initial lognormal distribution with central scale $M_c$ and variance $\sigma$ (black line). Its evolution after the 
accretion phase, assuming different cut-off redshifts, is shown in the blue and yellow lines, highlighting the appearance of a tail at high masses.

Finally, let us mention that also the PBH abundance in the universe is affected by baryonic accretion. By assuming a non-relativistic dominant component (whose energy density scales as the inverse of the volume), one can show 
that~\cite{DeLuca:2020fpg}
\begin{align}
	\label{fev}
	f_\text{\tiny PBH}(z)  &
	= \frac{\langle M(z)\rangle}{\langle 
		M(z_\text{\tiny i})\rangle(f^{-1}_\text{\tiny PBH}(z_\text{\tiny i})-1)+\langle M(z)\rangle}, \quad \langle M(z)\rangle =\int {\rm d} M M \psi (M, z)\,.
\end{align}
Because of accretion, $f_\text{\tiny PBH}(z)$ can be significantly larger than its value at formation, $f_\text{\tiny PBH}(z_\text{\tiny i})$, see Ref.~\cite{DeLuca:2020qqa} for details.

\subsection{PBH spin evolution}

If baryonic accretion is not efficient enough during the cosmic history, the spin of PBHs is natal. In the standard formation scenario during the radiation era, the dimensionless spin parameter $\chi = |\vec J|/G M^2$ at formation is of percent level~\cite{DeLuca:2019buf,Mirbabayi:2019uph,Harada:2020pzb}, although larger values can be achieved in other formation scenarios, see e.g.~\cite{Harada:2017fjm,Cotner:2017tir,Flores:2021tmc,Eroshenko:2021sez} and Chapter 10 for further details. 

On the other hand, in the case of efficient accretion, the formation of a thin disk leads to an efficient angular momentum transfer from the baryonic particles to the PBH, which could spin-up in the process. In other words, when the condition to form a thin accretion disk is satisfied, $\dot m\gtrsim 1$ (see discussion in Sec.~\ref{Sec1diskformation}),  mass accretion is accompanied by an increase of the PBH spin. For PBH binaries, the necessary condition $\dot m\gtrsim 1$ is even more easily fulfilled, such that
angular momentum transfer is more efficient~\cite{DeLuca:2020bjf,DeLuca:2020qqa}. 

Since the disk is expected to be located along the equatorial plane~\cite{Shakura:1972te,1973blho.conf..343N}, the PBH 
spin would align perpendicularly to the disk plane. In such a configuration, one can use a geodesic model to describe the angular-momentum growth~\cite{Bardeen:1972fi}. 
According to this model, the rate of variation of the Kerr parameter is related to the mass accretion rate $\dot M$ by~\cite{Bardeen:1972fi,Thorne:1974ve,DeLuca:2020bjf}
\begin{equation}
\dot \chi = \left( {\cal F} (\chi) - 2 \chi \right) \frac{\dot {M}}{M}\,,
\end{equation}
in terms of the quantity ${\cal F} (\chi) \equiv L(M,J)/M E(M,J)$, which is only a function of $\chi$ and depends on
\begin{equation}
E(M,J) = \sqrt{1- 2 \frac{GM}{3 r_\text{\tiny ISCO}}}
\quad 
\text{and}
\quad
L(M,J) = \frac{2 G M }{3 \sqrt{3} } \left( 1+ 2 \sqrt{ 3 \frac{r_\text{\tiny ISCO} }{G M }-2}\right)\,.
\end{equation}
In these equations, the innermost stable circular orbit (ISCO) radius reads
\begin{equation}
r_\text{\tiny ISCO}(M,J) = GM \left[ 3 + Z_2 - \sqrt{\left( 3-Z_1\right) \left( 3+Z_1+2 Z_2\right) } \right]\,, \label{ISCO}
\end{equation}
with $Z_1= 1+ \left( 1- \chi^2 \right) ^{1/3} \left[ \left( 1+\chi\right)^{1/3}+\left( 1-\chi\right)^{1/3} \right]$ and $Z_2= \sqrt{3 \chi^2 + Z_1^2}$\,.
One finds that, once the thin-disk hypothesis is satisfied,  the PBH spin is expected to grow over a typical accretion time scale $\tau_\text{\tiny acc}$, until it reaches 
the maximal value $\chi_\text{\tiny max}=0.998$, where radiation effects come into play and prevent reaching extremality~\cite{Thorne:1974ve}.

\begin{figure}[t]
\centering
\includegraphics[scale=.5]{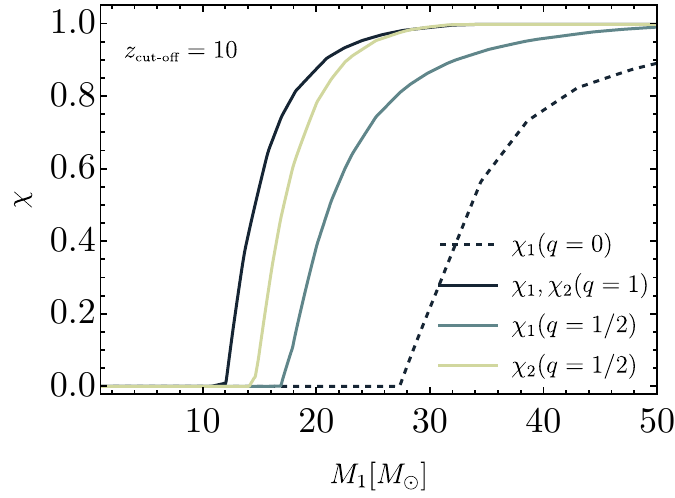}
\includegraphics[scale=.5]{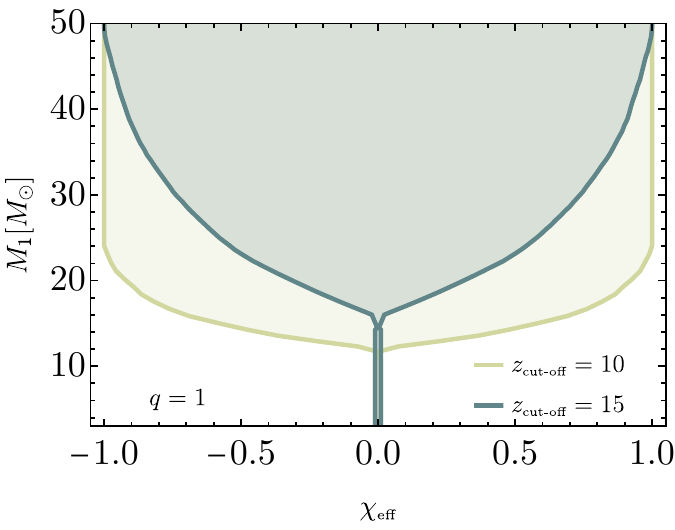}
\caption{Left panel: Behaviour of the PBH final spins $\chi$ in terms of the final masses $M$, assuming different values for the final mass ratio $q$. 
We have fixed the cut-off redshift to the representative value $z_\text{\tiny cut-off} = 10$. The dashed black line shows the evolution for a single PBH, while the solid lines hold for PBH binaries. 
Right panel: $99\%$ (3$\sigma$) confidence intervals for the effective spin parameter $\chi_\text{\tiny eff}$ for given final PBH masses $M_1 = M_2$, for different cut-off redshifts $z_\text{\tiny cut-off} = 15,10$. 
Figure adapted from Refs.~\cite{DeLuca:2020bjf,DeLuca:2020qqa}.}
\label{fig:spins}       
\end{figure}

In the left panel of Fig.~\ref{fig:spins}, we show the final PBH spins as a function  
of their final masses for different combinations of binary systems and assuming 
a cut-off reshift $z_\text{\tiny cut-off} = 10$. Similarly to Fig.~\ref{fig:masses},
the dashed black line shows the spin evolution of a single PBH, while the solid lines show the evolution of PBH spins in binaries. The plot shows a clear correlation between the final PBH mass and spin: low-mass PBHs are slowly spinning or non-spinning, whereas high-mass PBHs are rapidly spinning. The mass scale at which this continuous transition occurs depends on the cut-off redshift and on the system properties (mass ratio etc.). For example, equal mass binaries, characterised by a stronger accretion phase, transit to a maximally spinning configuration for lower masses compared to single PBHs or to PBHs in binaries with smaller mass ratios.
Finally, because of its relative larger accretion rate, the secondary binary component always spins more than the primary (see green and yellow lines in the plot).

The characteristic PBH mass-spin relations described in this Figure can be parametrically written using the expressions~\cite{Franciolini:2021xbq}
\begin{align}
\label{spin-mass-relation-fit}
    \chi_i(M_1,q, z_\text{\tiny cut-off})
    \approx 10^{-2}
    + {\rm Min}\left[ 
    0.988, 10^{f^b_{i}}
    \left( \frac{M_1}{M_\odot}-{f^ a_{i}}\right)^{f^c_{i}} 
    \theta\left(\frac{M_1}{M_\odot} -{ f^a_{i} }\right)
    \right],
\end{align}
where $i = \{1,2\}$ indicates the binary components, and the coefficients $f^{a,b,c}_{i}$ depend on the cut-off redshift $z_\text{\tiny cut-off}$ and final mass ratio $q$. Those functions are expanded as a polynomial series of the form~\cite{Franciolini:2021xbq}
\begin{align}
\label{fitmassspinrel}
f^{\alpha}_{i}(z_\text{\tiny cut-off},q) = 
\alpha_i^0+ 
\sum_{j=1}^3 \alpha_i^{z,j} \left( \frac{z_\text{\tiny cut-off}}{10}\right)^j
+
\sum_{j=1}^3\alpha_i^{q,j} q^j 
+ \sum_{j,k=1}^2 \alpha_i^{j,k} \left(\frac{z_\text{\tiny cut-off}}{10}\right)^j q^k,
\end{align}
with $\alpha = [a,b,c]$. The coefficients in the analytical expression are reported in Table~\ref{tabcoeff}.

\begin{table*}[t!]
\tiny
\renewcommand{\arraystretch}{1.2}
\caption{Numerical coefficients in the analytical fit described in Eq.~\eqref{fitmassspinrel} for the mass-spin relations of PBH binaries~\cite{Franciolini:2021xbq}.}
\label{tabcoeff}
\vspace{0cm}
\begin{tabularx}{1.\columnwidth}{
>{\centering}X>{\centering}X>{\centering}X
>{\centering}X>{\centering}X>{\centering}X
>{\centering}X>{\centering}X>{\centering}X
>{\centering}X>{\centering \arraybackslash}X
}
\midrule
\midrule
$a^0_1$ &  $a^{z,1}_1$ &$a^{z,2}_1$&$a^{z,3}_1$&
$a^{q,1}_1$ &$a^{q,2}_1$& $a^{q,3}_1$& $a^{1,1}_1$& $a^{2,1}_1$& $a^{1,2}_1$ &$a^{2,2}_1$
\\
$57.8531$ & $-66.8879$ &  $43.9529$ & $-5.46522$ &
$-56.4905$ & $39.4605$ & $-10.5127$ &
$37.4532$ &  $-17.5600$ & $-17.5899$ & $7.32670$ 
\\
\midrule 
$b^0_1$ & $b^{z,1}_1$ &$b^{z,2}_1$&$b^{z,3}_1$&
$b^{q,1}_1$ &$b^{q,2}_1$& $b^{q,3}_1$& $b^{1,1}_1$& $b^{2,1}_1$& $b^{1,2}_1$ &$b^{2,2}_1$
\\
$2.14680 $&$ -3.65483 $&$ 1.23732 $&$ -0.185276$&$ -1.59262$&$ -1.33445$&$ 0.940219$&$ 2.48367$&$ 0.0136971$&$ -0.313974$&$ -0.218091$
\\ 
\midrule
$c^0_1$ & $c^{z,1}_1$ &$c^{z,2}_1$&$c^{z,3}_1$&
$c^{q,1}_1$ &$c^{q,2}_1$& $c^{q,3}_1$& $c^{1,1}_1$& $c^{2,1}_1$& $c^{1,2}_1$ &$c^{2,2}_1$
\\
$0.441418$ & $-0.738179$ & $0.834177$ & $-0.175491$ & $-0.231674$ & $2.12451$ & $-0.787300$ & $-0.0461876$ & $-1.20687$ & $-0.234563$ & $ 0.477210$
\\ 
\midrule
$a^0_2$ & $a^{z,1}_2$ &$a^{z,2}_2$&$a^{z,3}_2$&
$a^{q,1}_2$ &$a^{q,2}_2$& $a^{q,3}_2$& $a^{1,1}_2$& $a^{2,1}_2$& $a^{1,2}_2$ &$a^{2,2}_2$
\\
$44.3220 $&$-72.7617$&$ 50.9837$&$ -8.27027$&$ 19.8378$&$ -33.8142$&$18.3605 $&$ -6.80676 $&$ 1.95003 $&$ 0.0581762$&$ -0.957243$
\\
\midrule
$b^0_2$ & $b^{z,1}_2$ &$b^{z,2}_2$&$b^{z,3}_2$&
$b^{q,1}_2$ &$b^{q,2}_2$& $b^{q,3}_2$& $b^{1,1}_2$& $b^{2,1}_2$& $b^{1,2}_2$ &$b^{2,2}_2$
\\
$3.65282$&$-6.94442$&$ 3.55860$&$-0.630911$&$ -0.474109$&$-0.199862$&$ 0.0523957$&$ 0.737077$&$ 0.0855668$&$ -0.178022$&$ -0.0212303$
\\ 
\midrule
$c^0_2$ & $c^{z,1}_2$ &$c^{z,2}_2$&$c^{z,3}_2$&
$c^{q,1}_2$ &$c^{q,2}_2$& $c^{q,3}_2$& $c^{1,1}_2$& $c^{2,1}_2$& $c^{1,2}_2$ &$c^{2,2}_2$
\\
$ -0.189439 $&$ 1.28502$&$ -0.587638$&$0.0864602
$&$ -0.905386 $&$ 1.25085
$&$ -0.346207 $&$ 0.158765$&$ -0.447706$&$ -0.0109165 $&$  0.0945026$
\\
\midrule
\midrule
\end{tabularx}
\label{tabfit}
\end{table*}

In the context of gravitational wave observations, a useful parameter entering the waveform is the effective spin parameter 
\begin{equation}
\chi_{\text{\tiny eff}}=\frac{\vec J_1 /M_1  + \vec J_2 /M_2 }{M_1+M_2}\cdot \hat L\,, \label{chieffdef}
\end{equation}
in terms of the direction of the orbital angular momentum $\hat L$. Its distribution is shown in the right panel of Fig.~\ref{fig:spins} where, following Refs.~\cite{DeLuca:2020bjf,DeLuca:2020qqa}, we have averaged over the angles between the total angular momentum and the individual PBH spins. The plot shows the parameter 
in terms of the final PBH mass $M_1$, for an equal mass binary $q = 1$, and for 
different values of the cut-off redshifts. For small PBH masses, when accretion is not efficient, the distributions peak around vanishing values, since the PBH spins kept their natal values. On the other hand, for PBHs heavier than about ${\mathcal O}(20)M_\odot$, accretion has been efficient enough in increasing the PBH spins near extremality and the distribution reaches large values. The  correlation between low (high) values of the masses and low (high) spins is manifest, with a sharp transition around $M\sim {\cal O} (20) M_\odot$ depending on the value of the cut-off redshift. In particular, larger cut-offs imply shorter accretion phases and therefore narrower distributions. This correlation provides a distinctive feature to distinguish PBHs from other compact objects, and could be used to identify PBHs among other BHs in gravitational wave data~\cite{Franciolini:2022iaa, Franciolini:2021xbq}.

\begin{acknowledgement}
V.DL. is supported by funds provided by the Center for Particle Cosmology at the University of Pennsylvania. He would like to thank his co-authors on this subject, G. Franciolini, P. Pani and A. Riotto.
N.B. acknowledges partial support from the National Science Foundation (NSF) under Grant No.~PHY-2112884. He would also like to thank his co-authors on this subject, V. Bosch-Ramon, L. Piga, M. Lucca, S. Matarrese, A. Raccanelli, and L. Verde.
\end{acknowledgement}

\bibliographystyle{unsrt}
\bibliography{Chapter_accretion.bib}

\end{document}